\documentclass[12pt,preprint]{aastex}

\slugcomment{}

\shorttitle{3-D modeling and distance of NGC 6369 }
\shortauthors{Monteiro et al.}

\received{2003 November 11}
\begin{document}

\title{3-D Photoionization Structure and Distances of Planetary Nebulae I.  
NGC 6369}

\author{Hektor Monteiro$^{1,2}$, Hugo E. Schwarz$^1$, Ruth Gruenwald$^2$}
\affil{1 Cerro Tololo Inter-American Observatory\altaffilmark{1}, Casilla 603, 
Colina El Pino S/N, La Serena, Chile}
\affil{2 Instituto de Astronomia, Geof\'isica e Ciencias Atmosf\'{e}ricas, 
S\~{a}o Paulo, Brasil}

\author{Steve Heathcote}
\affil{SOAR Telescope, Casilla 603, Colina El Pino S/N, 
La Serena, Chile}

\altaffiltext{1}{Cerro Tololo Inter-American Observatory, National Optical 
Astronomy Observatory, operated by the Association of Universities for
Research in Astronomy, Inc., under a cooperative agreement with the
National Science Foundation.}

\begin{abstract}
We present the results of mapping the planetary nebula NGC\,6369 using
multiple long slit spectra taken with the CTIO 1.5m telescope. We
create two dimensional emission line images from our spectra, and use
these to derive fluxes for 17 lines, the H$\alpha$/H$\beta$ extinction
map, the [SII] line ratio density map, and the [NII] temperature map
of the nebula. We use our photoionization code constrained by these
data to determine the distance, the ionizing star characteristics, and
show that a clumpy hour-glass shape is the most likely
three-dimensional structure for NGC\,6369. Note that our knowledge of
the nebular structure eliminates all uncertainties associated with
classical distance determinations, and our method can be applied to
{\it any spatially resolved emission line nebula}. We use the central
star, nebular emission line, and optical+IR luminosities to show that
NGC\,6369 is matter bound, as about 70\% of the Lyman continuum flux
escapes. Using evolutionary tracks from
\cite{B95} we derive a central star mass of about 0.65\,M$_{\odot}$.

\end{abstract}


\section{Introduction}

\begin{figure}
\includegraphics[scale=0.65]{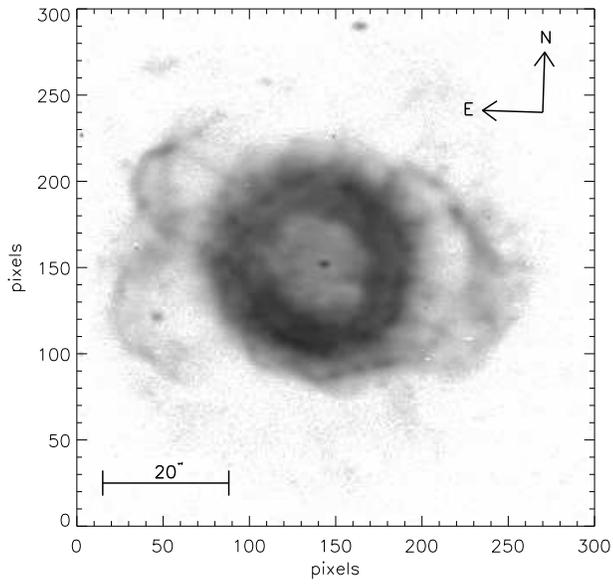}
\caption{Narrow band filter image of NGC\,6369 in 
H$\alpha $+{[}NII{]} from \cite{SCM92}. Note the bright ring and outer
faint ansae to the E and W of the object. N is up, E to the left, and
the plate scale is 0.26 ''/pix. \label{n2ima}}
\end{figure}

The planetary nebula (PN) NGC\,6369 -- shown in Fig.\,\ref{n2ima} in
the light of H$\alpha$\,+\,N[II]$\lambda$658.4nm -- is an object with
a complex morphology, consisting of a main bright annulus of diameter
40\arcsec, and fainter, curved outer structures on two sides in the
E-W direction. Deep narrow band H$\alpha$\, and
\,N[II]$\lambda$6584 images of NGC\,6369 were obtained by \cite{CSSP03}
in a survey to look for faint outer halos around PNe. Apart from
being deeper, their image is not significantly different from the one
shown in Fig.\,1, and no large faint halo was found.

The HST image of NGC\,6369 available at
http://heritage.stsci.edu/2002/25/index.html also shows the same main
features as our image but has additional resolved details, and tags
the H$\alpha$, [OIII]500.7nm, and [NII]658.4nm lines with the colors
red, blue and green respectively. In the section on observational
results we discuss this composite HST image in more detail.

\clearpage
\clearpage

The equatorial and galactic coordinates for NGC\,6369 are $\alpha$ =
17$^h$\,29$^m$\,20$^s$.40 and $\delta$ = 23$^o$\,45\arcmin\,37\arcsec.9
(ICRS2000), and l = 2.43$^o$ b = +5.85$^o$ respectively. The extreme distances
determined for NGC\,6369 are 0.33\,kpc obtained by
\cite{AGNR84}, and 2.00\,kpc obtained by \cite{GPG86}, out of a total of 
10 listed by \cite{AO92}, from which we obtain a mean value for the
distance to NGC\,6369 of 1039$\pm$ 538 pc. Note that these ten
distances were determined using several statistical methods, each with
their own systematic biases and assumptions, so that the average
probably is a reasonably unbiased estimator, within the (large) error.

The temperature of the central star obtained from the Zanstra method
by \cite{GP89} is $T(H)_\mathrm{Z}=67\,600~\mathrm{K}$. The star has
been classified as a WC4 star by \cite{TAS93}.

Spectra of NGC\,6369 were obtained by \cite{AKSJ1989}, using an IDS
(Image Dissector Scanner) as detector with double
4\,\arcsec~apertures, by \cite{AK87} using an ITS (Image Tube Scanner)
with a $2\arcsec\,\times\,10\arcsec$ slit, and by \cite{PPPP} using an
IDS and a combination of different apertures not bigger than
$13\arcsec\,\times\,13\arcsec$. An expansion velocity of $41.5$ km
s$^{-1}$ has been determined by \cite{MWF88}. The system radial
velocity was reported to be -106\,km.s$^{-1}$ by \cite{W53}, and
-101\,km.s$^{-1}$ by \cite{MWF88}.

\begin{figure}
\includegraphics[width=\columnwidth]{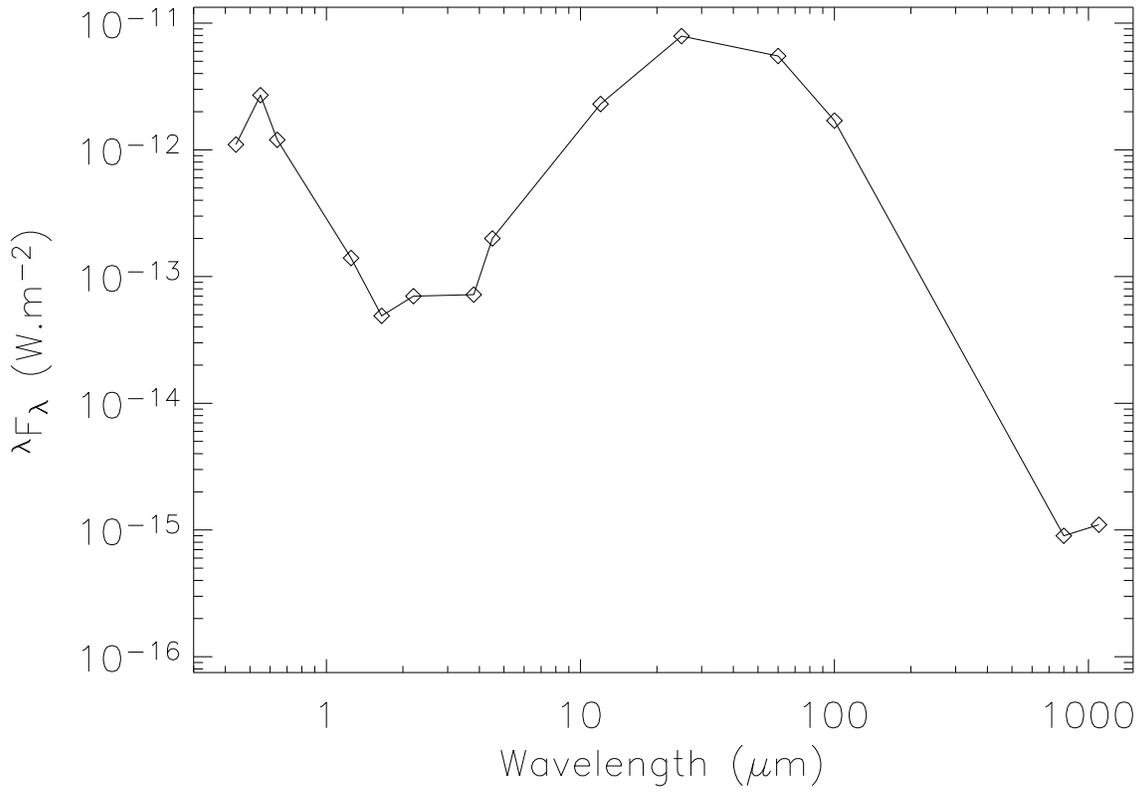}
\caption{The SED for NGC\,6369 between 0.44 and 1100\,$\mu$m using flux 
values from the literature (see text).\label{sed}}
\end{figure}

The spectral energy distribution (SED) of NGC\,6369 is shown in
Fig.\,\ref{sed} and has been computed using flux values from
\cite{AO92} and \cite{SK02} over a range of wavelengths from 
0.44 to 1100\,$\mu$m. Note that we plot $\lambda$\,F$_{\lambda}$ against 
wavelength and below we will use the F$_{\lambda}$ values to determine the 
observed optical+IR luminosity of NGC\,6369.

\clearpage
\clearpage

Planetary nebulae are generally classified according to their
appearance, and such classification is then used for studies of the
formation and evolution of these objects (for example
\citep{B87}). Such a classification, based on the observed two 
dimensional (2-D) brightness distribution in a given line or lines can
be quite misleading. This could be the case for NGC 6369, for which a
round morphology is evident from the observed images. Different 3-D
geometries can produce the same observed 2-D morphology as has been
shown by \cite{M00}. Furthermore, images produced by ions of different
ionization degree can be very different, due to the radiation transfer
within the nebulae. Only detailed modeling, which reproduces the
brightness intensity distribution of different lines, as well other
observed parameters, can provide more reliable information about the
3-D geometry of the gas distribution. The determination of the actual
3-D gas distribution in planetary nebulae is essential for
understanding their formation and evolution, as well as that of the
ionizing star. 

By determining the 3-D structure of nebulae, we eliminate the large
uncertainties that have plagued classical statistical distance
determination methods for over 5 decades. Assumptions about the
filling factor, constancy of ionized mass or diameter, mass-radius
relationships etc. are not needed here: we {\it know} what the
structure and ionized mass are, and can therefore determine distances
to much greater accuracy than before. In this paper we explain in detail how
we can determine this 3-D structure from long slit spectra and our
3-D photo-ionization model, and apply it to the case of NGC\,6369

In summary, we obtain the spatial structure of the object along with
its chemical composition, ionizing source temperature and luminosity,
mass, as well as an {\it independent} distance, in a self-consistent
manner. In \S2 we discuss the observations and basic reduction
procedures, as well as the details of the image reconstruction
technique used to obtain the line intensity maps. In
\S3 the results obtained from these maps are discussed: the reddening
correction of the images, total line fluxes and the computed
temperature and density maps. In \S4 we present the model results of
the 3-D photoionization code, and we discuss the derived quantities
Finally, in \S5 we give our conclusions.

\section{Observations and Mapping}

\subsection{Observations and general reduction}

Observations were made using the CTIO 1.5\,m Ritchey-Chr\'etien
telescope with the RC Spectrograph on the nights of 12 and 13th June,
2002. We used a grating with 600\,l/mm blazed at 600\,nm giving a
spectral resolution of 0.65\,nm per pixel and a plate scale of
1.3\arcsec/pixel with a slit width of 4\arcsec.  The spectral coverage
obtained with this configuration was approximately 450\,nm to
700\,nm. We took three 1200s exposures at each slit position for ease
of cosmic hit removal etc.. For details of the instrument and
telescope see http://www.ctio.noao.edu and click on ``Optical
Spectrographs'', then on ``1.5M RC spectrograph''.

By taking sets of exposures at several parallel long-slit positions
across the nebula, we obtained line intensity profiles for each
slit. These profiles were then combined to create emission line images
of the nebula with a spatial resolution of about
$4\arcsec\,\times\,4\arcsec$, in a way similar to radio mapping. This
method has been applied before for the PN NGC\,3132 as discussed in
\cite{M02}. A total of 10 positions were observed, moving the slit in 4\arcsec
steps in the N-S direction between exposures. The third position
observed from south to north (east is to the left, and north is up in
all the figures) was not observed due to pointing problems. For this
position an average of the two adjacent exposures was adopted.

The seeing conditions during the observing run were 1.5\arcsec or
better, obtained from the seeing monitor at CTIO. In both nights the
atmospheric conditions were photometric.

The individual slit spectra were reduced using standard procedures for
long-slit spectroscopy, using IRAF reduction packages.  The wavelength
calibration was applied to the two dimensional images using the tasks
FITCOORDS and TRANSFORM after re-identifying the arc spectra in many
spatial positions. This procedure helps to remove distortions in the
image that may be present due to slit curvature and misalignment with
respect to the CCD. The final corrected images show displacements of
the order of one pixel, which is the precision limit of this method.
An additional fine correction for slit misalignment was made using the
H$\alpha $ and H$\beta $ profiles for each exposure. Using IDL the
images were re-dimensioned to 100 times their original size. The
normalized H$\alpha $ and H$\beta $ profiles were then matched and the
final result re-dimensioned to original values. This procedure yields
the precise alignment necessary for calculation of diagnostic line
intensity ratios. Minor shifts of the order of one pixel can introduce
considerable errors in line ratios, if this method is not applied. The
error introduced in line ratios by misalignments can reach 4\% on sharp
edges of the brightness profile.

For the flux calibrations two standard stars, LTT\,7379 and LTT\,9239
were observed on both nights. The final sensitivity function obtained
from these stars, using the IRAF task SENSFUNC, has an RMS of 0.02,
indicating the good photometric quality of the nights.  Based on
previous similar spectrophotometric observations made with the INT on
La Palma, we compute the {\it random} error on individual line flux
calibrations to be 3\%, while the overall possible {\it systematic}
uncertainty in the stellar flux is about 5\%.

\subsection{Image reconstruction}

The final emission line images were obtained from the spatially
integrated flux profile in each slit. To obtain the slit profile of
integrated fluxes in a given transition, a Gaussian function was
fitted to each point along the spatial direction using IDL
routines. The errors adopted for the fits were the standard deviations
calculated by the fitting procedure. The profiles for each slit
position were then combined and interpolated using a cubic convolution
algorithm \citep{PS83,RM74}, in order to reconstruct a 2-D image of the
nebula for that transition. Integration of this final image provides
the total observed flux for the nebula, for each line.  Note that this
procedure gives better estimates of the total flux and relative line
intensities than observing just one slit position or aperture, since
it takes into account the entire nebula.

Signal to noise images were also obtained using the fitted profiles by
subtracting the fitted profile from the original section and analyzing
the remaining noise. This is a combination of all noise sources in the
image (photon statistics, noise introduced by the data reduction
procedures, etc.). These errors were then combined with the fitting
errors mentioned above, using the usual error propagation expressions
to make the final signal to noise image. We then quadratically added
the 3\% random calibration error and the de-reddening error computed
from the H$\alpha$/H$\beta$ reddening map (see \S3.1), to arrive at
the overall error for each line flux. This procedure provides precise
error estimates for the total fluxes and diagnostic line ratio
calculations.

\section{Observational results}

\begin{figure*}
\includegraphics[]{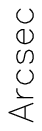}
\caption{Reddening corrected, reconstructed emission line contour plots 
for NGC6369. The peak intensity in each frame is normalized to unity,
and there are 7 contour steps between the maximum (black) and zero
(white). N is up, E to the left.\label{maps_full} }
\end{figure*}

Images were created for all 17 lines detected with sufficient
signal-to-noise ratio. In Fig.\,\ref{maps_full} the images for the
most important lines are shown. The important [OIII]436.3nm line fell
outside our available spectral range and could therefore not be
included in our analysis. The images were corrected for reddening as
described in the following section. The extracted total line
intensities with their fractional total errors are given in Table~1.

Note that all main features are reproduced in these reconstructed line
images.  The outer ansae only show up in the [NII]685.4nm line, and
are extremely faint or invisible in all other lines, as they are in
the HST original images shown at
http://heritage.stsci.edu/2002/25/original.html The [OIII]500.7nm
emission comes mainly from the central part of the bright ring-like
structure, while H$\alpha$ comes from further out, and the outermost
part is dominated by [NII]685.4nm. Clearly, we do not reproduce the
fine detail visible in the HST image because of our effective spatial
resolution of 4'' in the observational data. However, the expected
ionization stratification for the different ions and the overall
nebular structure are clearly seen in our images.

\subsection{Reddening correction and total fluxes}

\begin{figure}
\includegraphics[width=\columnwidth]{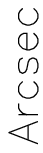}
\caption{H$\alpha $/H$\beta $ ratio map with contour overlay of H$\alpha $
(Image is cut for S/N lower than 10). N is up, E to the left. \label{ha/hb}}
\end{figure}

The reconstructed images for each line were corrected for reddening
using the H$\alpha $/H$\beta $ ratio map shown in Fig.\,\ref{ha/hb}.
The logarithmic correction constant was calculated pixel by pixel
using the theoretical value of H$\alpha $/H$\beta $=2.87 (\citep{O89})
and the reddening curve of \cite{S79}. We investigated the effect of
differential atmospheric refraction on this ratio map. From the
airmasses of our observed positions and the values given by \cite{F82}
we computed a correction which we applied to our data. Since we used
wide slits (4\arcsec) and the object is extended, the effect was
small, but not negligible in the steep gradients near the bright ring
structure. The average error due to this effect is about 2\% in the
high signal to noise areas and about 20\% in the low signal to noise
areas. The nett effect on the final calculated relative total fluxes
is about 0.2\% for strong lines and of 5\% for weak ones, well within
the other observed uncertainties.

\begin{figure}
\includegraphics[width=\columnwidth]{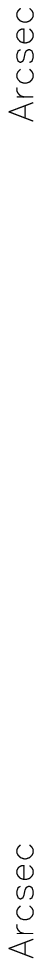}
\caption{Comparison of [NII] image before and after reddening correction.
N is up, E to the left.
\label{n2corrected}}
\end{figure}

In Fig.\,\ref{n2corrected} we show the {[}NII{]}658.4nm image as an
example before and after the reddening and differential refraction
corrections.

\clearpage

\clearpage
\clearpage

The reddening correction was also obtained from the spatially
integrated fluxes calculated from the images. The H$\alpha
$ flux was divided by the H$\beta $ flux and a single value
for the logarithmic extinction constant was obtained for the whole
nebula. This value was then used to correct the total line
intensities. The results for the line intensities were compared with
those obtained from the first method and showed no differences to
within the calculated uncertainties.

The resulting emission line fluxes relative to H$\beta$ and their
corresponding total errors are shown in Table\,\ref{line fluxes}. All
total fluxes were obtained by integrating the reddening corrected
(pixel by pixel) images. The value we obtained for the total
de-reddened H$\beta$ flux was $F_{H\beta
}\,=\,7.2\,\times\,10^{-13}W\,m^{-2}$ and the uncorrected flux
5.9$\,\times\,10^{-15}W\,m^{-2}$ corresponding to a value of
E(B-V)\,=\,1.4 magnitudes.

\clearpage
\begin{table}

\centering

\caption{Line fluxes relative to H$\beta$ \label{line fluxes}}

\begin{tabular}{lccc}
\hline
{ Line}& { Flux}& {Dered. Flux}& {Error\,(\%)}\\
\hline\hline

	  {HeII$\lambda$468.6} & {0.011} & {0.014} & {19}\\
	  {[OIII]$\lambda$495.9} & {4.59} & {4.11} & {7}\\
	  {[OIII]$\lambda$500.7} & {14.60} & {12.41} & {7}\\
	  {[CIII]$\lambda$551.8} & {0.01} & {0.01} & {25}\\
	  {[CIII]$\lambda$553.8} & {0.01} & {0.01} & {22}\\
	  {[NII]$\lambda$575.5} & {0.04} & {0.02} & {17}\\
	  {HeI$\lambda$587.6} & {0.47} & {0.16} & {8}\\
	  {[OI]$\lambda$630.0} & {0.22} & {0.05} & {8}\\
          {[SIII]$\lambda$631.1}& {0.05} & {0.014} & {13}\\
	  {[OI]$\lambda$636.3} & {0.07} & {0.02} & {13}\\
          {[NII]$\lambda$654.9} & {1.20} & {0.24} & {7}\\
	  {$H\alpha\lambda$656.3} & {13.80} & {2.87} & {7}\\
	  {[NII]$\lambda$658.4} & {3.70} & {0.73} & {7}\\
	  {HeI$\lambda$667.8} & {0.22} & {0.04} & {9}\\
	  {[SII]$\lambda$671.7} & {0.25} & {0.04} & {10}\\
	  {[SII]$\lambda$673.1} & {0.37} & {0.07} & {9}\\
	 
\hline

\end{tabular}
\end{table}
\clearpage

\subsection{Gas density and temperature}

\begin{figure}
\includegraphics[width=\columnwidth]{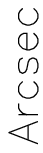}
\caption{Density map obtained from the de-reddened {[}SII{]}671.7,673.1 
line ratio using the method of \cite{D87}. N is up, E to the left. \label{dens_map}}
\end{figure}

We calculated the density and temperature maps from the reddening
corrected maps of the {[}SII{]} and {[}NII{]} lines respectively. The
expression relating the sulfur doublet line intensity ratios to the
gas density is from \cite{D87} which is also used in the IRAF package {\it
temden}. The density map is shown in Fig.\,\ref{dens_map}. For the
temperature map we used the expression given by \cite{M84} as this
allowed us to take the density variations over the nebula into
account, which we did by using our density map. The difference between
using this method and one that assumes constant density is small
--since the temperature is only weakly dependent on the density-- but
systematic.

\begin{figure}
\includegraphics[width=\columnwidth]{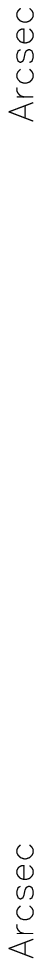}
\caption{Temperature map obtained from the de-reddened [NII] line ratio (A), 
and the difference between this map and the one obtained using constant 
density (B). Note that the maximum difference is less than 500\,K. N is up, 
E to the left.
\label{temp_comp}}
\end{figure}

For the temperature maps, correction for slit misalignment was carried
out as done for the H$\alpha $/H$\beta $ maps discussed in section
2.1. This correction was not necessary for the density maps since the
small wavelength separation of the two lines generates very little
deviation. The temperature map obtained after de-reddening is shown in
Fig.\,\ref{temp_comp}(A). We also show the difference between the
temperature map obtained using constant density and our map in
Fig.\,\ref{temp_comp}(B). Note that the differences are less than 500\,K.
The images are clipped for data values with S/N lower than 10 for
visual clarity.

\clearpage

\clearpage
\clearpage

\subsection{Temperature of the central star}

With the total fluxes corrected for interstellar reddening, the central 
star temperature can be calculated. For this, we apply the method described by
\cite{HS66}, using the observed fluxes presented in \S3.1 and the
central star flux $F_{4793\AA}= 1.20\,\times\,10^{-15}\,erg\,cm^{-2}\,
s^{-1}$ obtained from \cite{GP88}. It is important to note that this
method assumes spherical symmetry and a black-body spectral energy
distribution for the ionizing star. This is not necessarily true as
has been shown by many authors (for recent work see \cite{RDDW00})
Here we use a blackbody spectrum modified by the major H and He
absorption edges, and below we show by comparison with stellar
evolution models from \cite{B95} that this is a reasonable method.

From the H$\beta $ and $HeII \lambda 468.6$ line fluxes, we obtain,
respectively, $T_\mathrm{Z}(H)= 70000~\mathrm{K}$ and
$T_\mathrm{Z}(HeII)= 106000~\mathrm{K}$. The value obtained for
$T_\mathrm{Z}(H)$ with our fluxes differs slightly from the one
obtained by \cite{GP89} but is within the expected (large)
uncertainties of the method. These values are more reliable than
previously calculated ones since we have precise integrated fluxes for
the whole nebula.

The Zanstra temperature for H and HeII are not necessarily equal, and
do not correspond to the effective temperature of the central star if
the nebula is not optically thick in all directions
\citep{GV00}. NGC\,6369 is optically thin for radiation short-ward of
91.2\,nm (Lyman continuum) since a rough estimate of the total
luminosity in the optical lines is $\approx$\,14\% of that shortward
of 91.2\,nm. Most of the UV escapes the nebula, and therefore we
expect the T(H) do be lower than the T(He), as indeed found above.

\section{Photoionization Models}

The photoionization code applied here was described in detail by
\cite{GVB97}. It uses a cube divided into cells, each
having a given density. This allows arbitrary density distributions to
be studied. The input parameters are the ionization source
characteristics (luminosity, spectrum, and temperature), element
abundances, density distribution, and the distance to
the object. The conditions are assumed uniform within each cell where
the code calculates the temperature and ionic fractions. These values
are used to obtain emission line emissivities for each cell.

The final data cube can be oriented in order to reproduce the observed
morphology. The orientation of the structure is thus also
determined. The line intensities and other relevant quantities are
then obtained after projection onto the plane of the sky. Note that
because we lack velocity information, we cannot specify if an ansa is
in front of or behind the main nebula.

\begin{figure}
\includegraphics[width=\columnwidth]{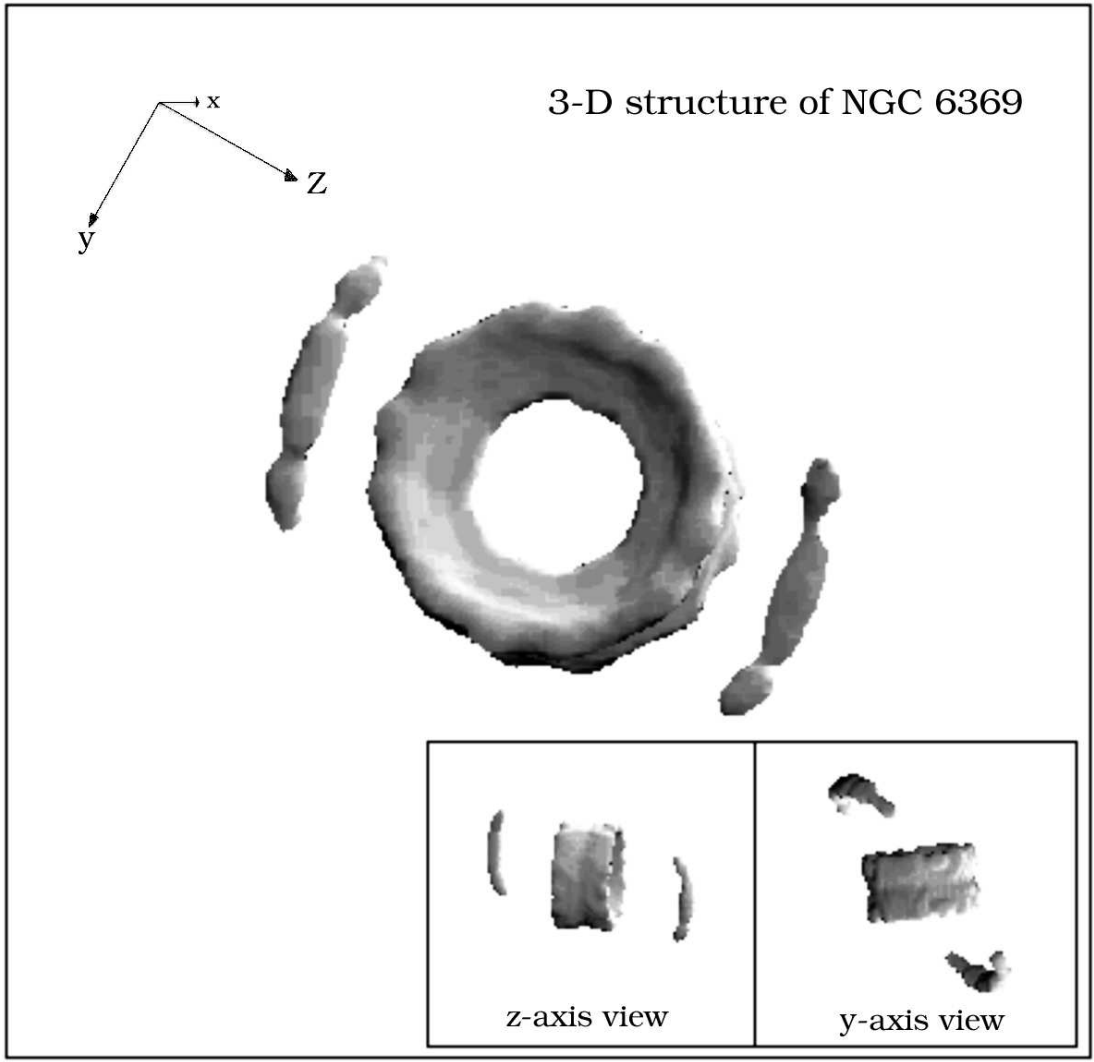}
\caption{Iso-density plot of the 3-D structure used for the NGC\,6369 
model calculations. In the main image N is up, E to the left. The
small insets view the structure along the other two axes, allowing the
3-D structure and orientation of the central object and the ansae to be
clearly seen.\label{strut3D}}
\end{figure}

The structure adopted for NGC 6369 is shown in Fig.\,\ref{strut3D} as
it is oriented relative to the observer. The choice of the structure
was based on the density map obtained from the observations. As shown
in \cite{M00}, closed shells were not able to reproduce the decrease
in the density profile in the central regions of NGC\,3132. For the
same reason, we adopt an hour-glass shape for the main structure of
NGC\,6369, as its density map shows the same decrease in the central
regions. We include random fluctuations in the density grid to
simulate the clumpiness observed in images, as well as a density gradient
along the main axis of the bipolar structure. The fluctuations were
set to vary between the maximum and the minimum densities measured
from the observational data. The rotation angles relative to the x, y
and z axis are 70$^o$, 10$^o$, and 0$^o$ respectively, with the
symmetry axis of the main structure being x. The model resolution of
$100^3$ cells was limited by computer memory.

The ionizing spectrum adopted for the central source is a modified
blackbody with a break at 54.4eV. This was necessary to fit both the
[OIII] and HeII line intensities, the last one being particularly
sensitive to the depth of the break. This break is also present in
the more precise theoretical atmospheric models presented by Rauch et
al. (2003). The cutoff at 54.4\,eV is one of the most prominent
features that differentiates this spectrum from that of a blackbody,
followed by the 13.6\,eV hydrogen cutoff and absorption lines. The effective
temperature and luminosity of this modified blackbody are listed in
Table\,\ref{mod-res}.

\clearpage
\clearpage

\subsection{Model Results}

\begin{figure*}[!ht]
\includegraphics[]{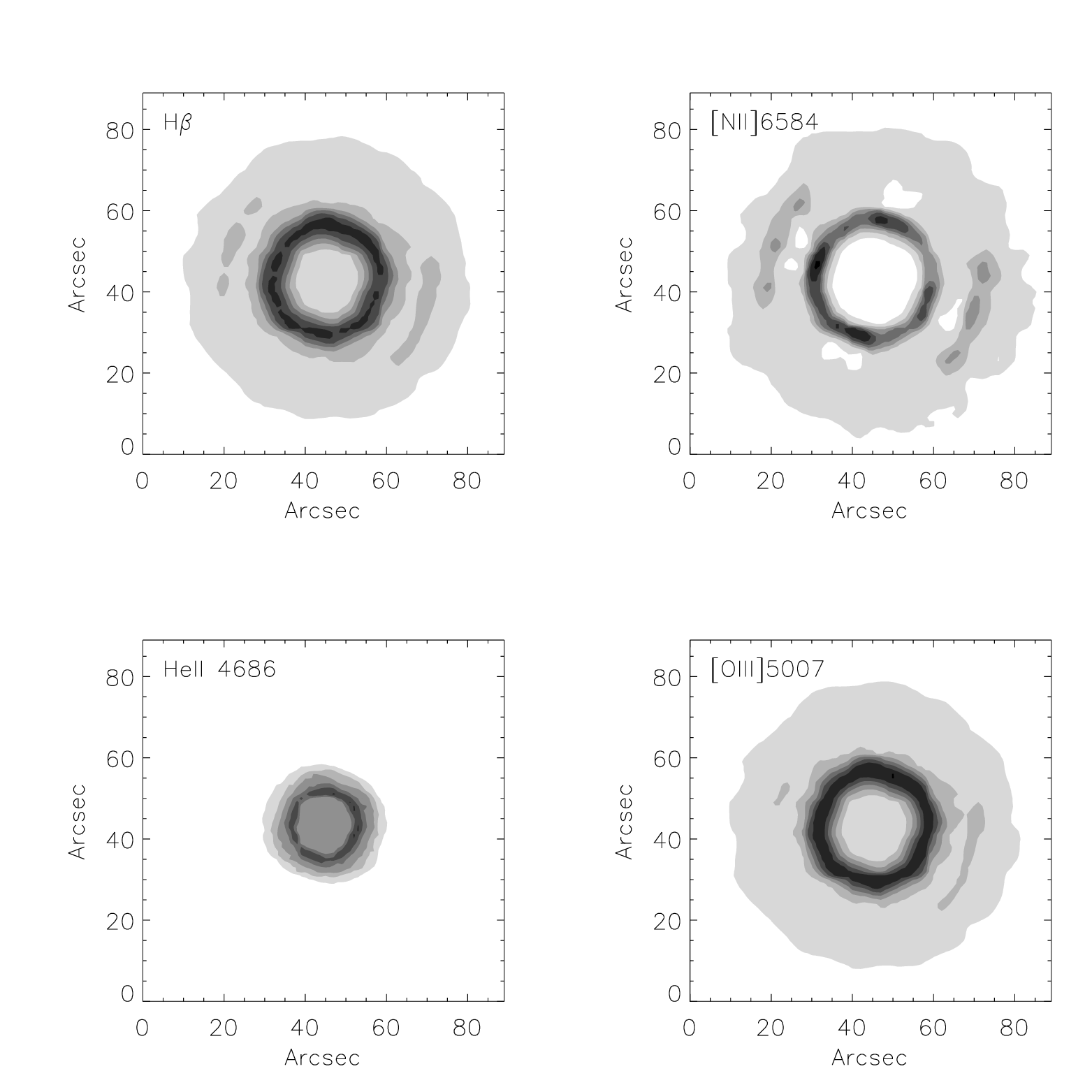}
\caption{Projected line images obtained with the photoionization 
model. N is up, E to the left.\label{ims-mod}}
\end{figure*}

We present here the main results obtained with the photoionization
model. The total line intensities are given in Table\,\ref{mod-res},
as well as the fitted abundances and ionizing star
characteristics. Projected line images for the most important
transitions are shown in Fig.\,\ref{ims-mod}.

\begin{figure}
\includegraphics[width=\columnwidth]{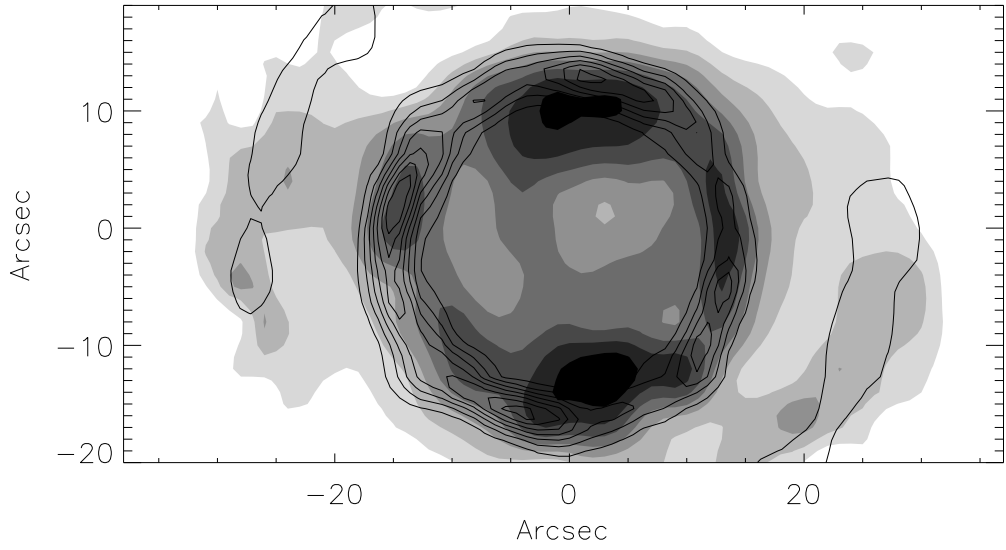}
\caption{Contours of the model [NII] image over the observed image for d=1550 
pc. N is up, E to the left. \label{mod-comp}}
\end{figure}

The model image size is fitted to the observed size for the
[NII]$\lambda658.4\,nm$ line, as well as constrained by the absolute
$H\beta$ flux. The distance of 1550\,pc we determine is also
constrained by fitting all the line fluxes and the shape of the nebula
simultaneously, and for this reason is fundamentally different from
classical distance determination methods, as it needs no assumptions
to be made. One cannot increase the distance, and simply make the
nebula larger to fit the optical image, as the line ratios would
change significantly. 3-D ionization models such as this one cannot
simply be scaled, thus placing much more stringent constraints on all
derived parameters, including the distance. Note that this model
distance is within one adjusted standard deviation of the average of
all the observationally determined distances given in
Sec.\,1. Fig.\,\ref{mod-comp} shows the observed [NII] image with the
corresponding model image contours overlaid for the obtained distance.
   
\clearpage
\clearpage

\begin{figure}
\includegraphics[width=\columnwidth]{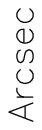}
\caption{The [SII] line ratio obtained with the model. N is up, E to the left.\label{s2ratio}}
\end{figure}

\begin{table}
\begin{center}
\caption{Observed and model line fluxes and model central star 
properties.\label{mod-res}}
\begin{tabular}{lll}\hline

	&   {Observed}
	&   {Model}
\\\hline\hline
	
	&
	&
\\
	    {$T_*$ (K)}	&   {70kK-106kK}	&   {91kK}\\
	    {$L_*/L_{\odot}$}	&   {-}	&   {8100}\\
	    {Density}	&   {1000-5400}	&   {300-5400}\\
	    {He/H}	& {-}	& {$1.14\times 10^{-1}$}\\
	    {C/H}	& {-}	& {$3.3\times 10^{-4}$}\\
	    {N/H}	& {-}	& {$1.0\times 10^{-4}$}\\
	    {O/H}	& {-}	& {$6.1\times 10^{-4}$}\\
	    {Ne/H}	& {-}	& {$6.5\times 10^{-5}$}\\
	    {S/H}	& {-}	& {$7.5\times 10^{-6}$}\\
	    {log($H\beta$)}	&   {-12.14}	&   {-12.16}\\
	    {[NeIII]386.8$^{a}$}&   {0.9}&   {0.9}\\
	    {[NII]658.4}  &   {0.73}	&   {0.78}\\
	    {[OIII]500.7}&   {12.41}	&   {12.33}\\
	    {HeII468.6}	&   {0.014}	&   {0.013}\\
	    {HeI587.6}	&   {0.163}	&   {0.169}\\
	    {[OI]630.0}	&   {0.052}	&   {0.048}\\
	    {[SII]671.7}&   {0.044}	& {0.047}\\	
	    {[SII]673.1}&   {0.066}	& {0.063}\\	
	&
	&

\\\hline

\end{tabular}

$^{a}$ Value obtained from \cite{PSM01}
\end{center}
\end{table}
\clearpage
The [SII] ratio map obtained with the model is shown in Fig.\,\ref{s2ratio}. 

\clearpage
\clearpage

\section{Discussion and conclusions}

We have presented spectrophotometric maps of NGC\,6369. These maps 
provided spatially resolved information for many emission lines and
precise total fluxes for the whole nebula. The images produced with
this technique were used to study the nebula with the usual diagnostic
ratios. Each image was corrected for reddening pixel by pixel using
the H$\alpha $/H$\beta$ image. This correction lead to some
significant differences between the corrected and uncorrected images,
as was shown in Figs.\,5 \& 7.

The H$\alpha $/H$\beta$ map shows some interesting features. In
Fig.\,4 we show that the extinction is not uniform across the nebula,
as it varies between 6 and 18. The structure follows the main nebular
morphology, which could indicate that dust and neutral material are
present. This is to be expected since the nebula shows a very clumpy
structure, which may cause shadowing within the gas, allowing for the
survival of grains and neutral material. The prominent features on
either side of the nebula do not show significant differences in
extinction when compared to the main region.

The temperature map derived from the observed integrated line of sight
intensities showed some structure but due to the low signal to noise
ratio, this should be treated with caution. Within the errors and
resolution of our observations the temperature can be considered to be
constant across the nebula.

We also show that the density map indicates a decrease in density
for the central regions. This decrease, as has been discussed by
\cite{M00}, is not compatible with a closed shell structure. Based on
this map we propose an hour-glass structure for the main nebula, which
has reproduced all the observational features of NGC\,6369. This
indicates that the one needs to use the [SII] ratio in addition to images
to distiguish between open and closed structures.

The position of the outer condensations or ansae --which are off-set
by 30$^o$ from the main nebular symmetry axis-- was obtained by
matching model images with the observed line images
(Fig.\,\ref{maps_full}) as well as matching the model [SII] density
map with the observed map (Fig.\,\ref{dens_map}). The presence of
these condensations can be accounted for by earlier ejection of matter
and precession can account for their deviation from the symmetry
axis. Many authors have dealt with these issues; for more complete
discussions and models see for example
\cite{CFJ96}, \cite{CFLJ95}, \cite{G97}, and \cite{LP97}.

Some of the parameters we have derived (and use as a sanity check)
are:

The total luminosity of the observed lines,
L$_l$\,=\,1150\,L$_{\odot}$; that derived from the SED by integrating
the F$_{\lambda}$ curve, L$_o$\,=\,7.1$\,\times$\,10$^{-4}$\,d$^2$
where d is the distance to NGC\,6369 in pc, resulting in
L$_o$\,=\,1700\,L$_{\odot}$ . Correcting this value using the method
of Myers et al. (1987) we obtain 2550\,L$_{\odot}$. The luminosity of
the central star is L$_{tot}$\,=\,8100\,L$_{\odot}$ so that the ratios
of the line and optical+IR luminosities to the total luminosity are
respectively: L$_l$/L$_{tot}$\,=\,0.14 and
L$_o$/L$_{tot}$\,=\,0.3. Assuming that the absorbed flux is
re-radiated in the IR, mainly the IRAS bands, the integrated
optical+IR luminosity indicates that about 70\% of the UV flux (Lyman
continuum) escapes from the nebula. This is not implausible as the
nebula is open at it's ``poles'' and is clumpy, allowing the radiation
to pass through the many ``holes''.

\begin{figure}
\includegraphics[width=\columnwidth]{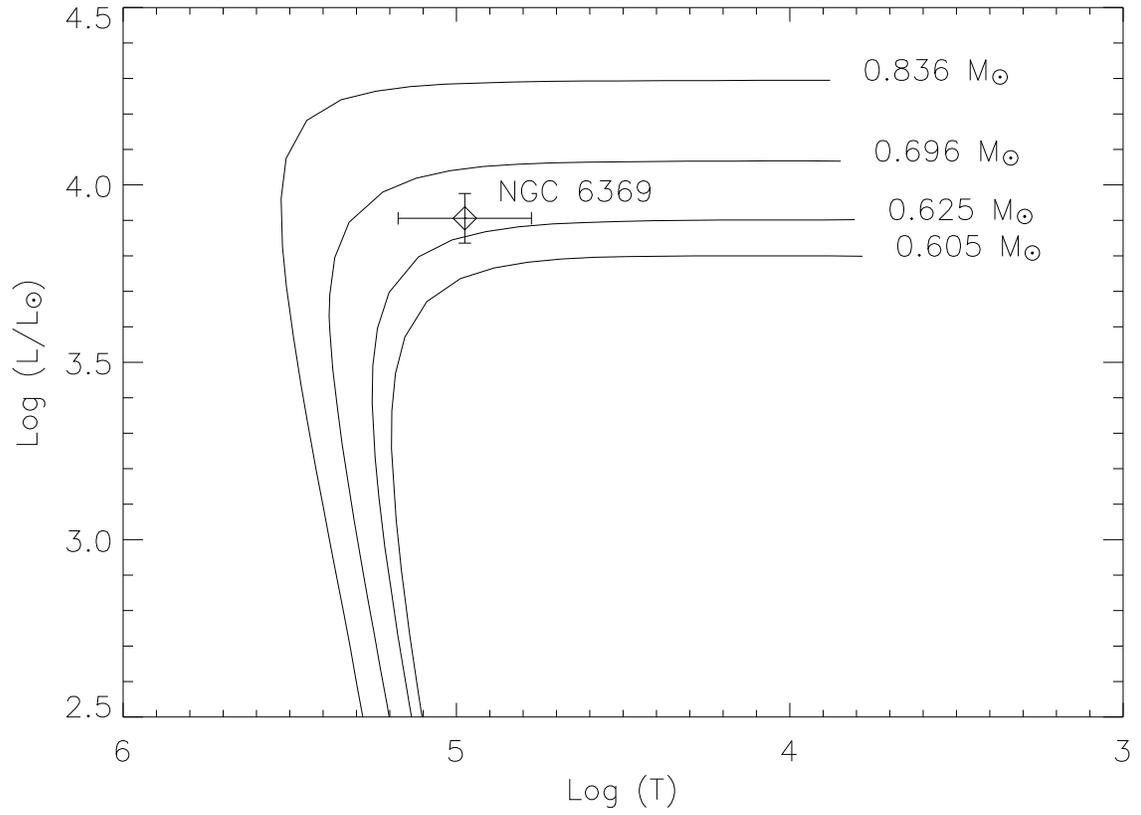}
\caption{Comparison of model temperature and luminosity obtained for the 
central star with model tracks calculated by \cite{B95}.\label{evol_track}}
\end{figure}

From the input matter distribution and abundances used in the model,
we calculate the mass of the nebular ionized gas to be
$M_{neb}\,=\,1.8\,M_{\odot}$. If we now use our values of luminosity
and temperature for the central star and compare them with the
evolutionary models of
\cite{B95} we can obtain the mass of the central core. Using this
procedure we can see from Fig.\,\ref{evol_track} that the best fitting
track for our data corresponds to a mass somewhat higher than
$M_{core}\,=\,0.625 M_{\odot}$, say, $\approx\,0.65\,M_{\odot}$, and
to an initial mass of $M_{0}=3 M_{\odot}$. If we sum our nebular and
core masses, we obtain $M_{core}\,+\,M_{neb}\,=\,2.4\,M_{\odot}$
approximately. If we use the line intensity errors as a measure of the
goodness of the model results and combine the luminosity and
temperature errors we obtain approximately 20-25\% for the uncertainty
in $M_{core}\,+\,M_{neb}$, placing the obtained initial stellar mass
within the uncertainty of our determination. The value for $M_{neb}$
calculated from the model input density is a lower limit estimate of
the total nebular mass as at least some material is present in the
form of dust, as indicated qualitatively by our extinction map
structure. Additionally, there may be neutral gas for wich we have no
constraints.

Note that typical parameters for a WC4 PN central star from \cite{A64}
are T\,=\,90kK, and R\,=\,0.4\,R$_{\odot}$, very close to our derived
values of 91000\,K and 0.4\,R$_{\odot}$. Using
g\,=\,G\,.\,M\,/\,R$^2$ in CGS units, we obtain log(g)\,=\,5.1, which
is much lower than the value of 5.8 derived from the depth of our He
break. This is due to the fact that the central star has an extended
atmosphere, again showing the consistency of the model results.

\clearpage
\clearpage

Using our photoionization code and the proposed structure we obtained
a complete 3-D model for the NGC\,6369. The fitted model line
intensities show excellent agreement with the observed values. The
obtained distance of d=1550 pc, is well within the range present in
the literature obtained from different methods. The model temperature
for the ionizing star is similar to the Zanstra (He) value discussed
in \S3.3. The temperature and luminosity values obtained from the
model as well as the total nebular plus central star mass show good
agreement with the stellar evolution models of \cite{B95}.

Using multiple long slit spectroscopy, we can determine accurate
distances, 3-D structures, abundances, ionized masses, central star
masses, luminosities, and temperatures to any spatially resolved
emission line nebula, assuming there are no strong shocks or extreme
morphologies involved.

\begin{acknowledgements}
      HM acknowledges support from Fapesp grant 00/03126-5, and 
      NOAO's Science Fund.
\end{acknowledgements}

\end{document}